\documentclass[a4paper,fleqn,usenatbib]{mnras}

\usepackage[T1]{fontenc}
\usepackage{ae,aecompl}

\usepackage{graphicx}	
\usepackage{amsmath}	
\usepackage{amssymb}	

\title[LMC star clusters]{A likely runaway star cluster in the outer disc of the
Large Magellanic Cloud}

\author[Piatti et al.]{
Andr\'es E. Piatti$^{1,2}$\thanks{E-mail: andres@oac.unc.edu.ar}, Ricardo Salinas$^{3}$
and Eva K. Grebel$^{4}$\\
$^{1}$Consejo Nacional de Investigaciones Cient\'{\i}ficas y T\'ecnicas, Av. Rivadavia 1917, 
C1033AAJ, Buenos Aires, Argentina\\
$^{2}$Observatorio Astron\'omico, Universidad Nacional de C\'ordoba, Laprida 854, 5000, 
C\'ordoba, Argentina\\
$^{3}$Gemini Observatory, Casilla 603, La Serena, Chile\\
$^{4}$Astronomisches Rechen-Institut, Zentrum f\"ur Astronomie der Universit\"at Heidelberg, 
M\"onchhofstr. 12-14, 69120 Heidelberg, Germany\\
}

\date{Accepted XXX. Received YYY; in original form ZZZ}

\pubyear{2018}

\begin{document}
\label{firstpage}
\pagerange{\pageref{firstpage}--\pageref{lastpage}}
\maketitle

\begin{abstract}
We present results from photometric and spectroscopic data obtained
with SOAR and Gemini observatory facilities in the field of a recently 
discovered star cluster. The cluster, projected towards the Eastern side of 
the outer disc of the Large Magellanic Cloud (LMC), was originally placed nearly 
10 kpc behind the LMC with an age and metallicity typical of the innermost
LMC star cluster population. We assigned radial velocity (RV) memberships to stars
observed spectroscopically, and derived the cluster age and distance from
theoretical isochrone fitting to the cluster colour-magnitude diagram.
The new object turned out to be a 0.9 Gyr old outer LMC disc cluster,
which possibly reached the present position after being scattered from the
innermost LMC regions where it might have been born. We arrived at this 
conclusion by examining the spatial distribution of LMC star 
clusters of similar age, by comparing the derived spectroscopic metallicity
with that expected for an outside-in galaxy formation scenario, by considering
the cluster
internal dynamical stage as inferred from its derived structural parameters and
by estimating the circular velocity of a disc that rotates with the corresponding 
star cluster radial velocity  at the cluster's deprojected distance, which resulted to be nearly 
60 per cent higher than that of most of the outer LMC disc clusters. 
\end{abstract}




\section{Introduction}

Recently, \citet{p16} discovered a star cluster towards the
Eastern part of the Large Magellanic Cloud (LMC) outer disc that appeared 
to be located behind the LMC at the Small Magellanic Cloud (SMC)'s distance. 
He speculated about the possibility of its being the first discovered cluster that 
was born in the LMC and soon ejected into the intergalactic space during the 
recent Milky Way/Magellanic Clouds (MW/MCs) interaction 
\citep{kallivayaliletal13,casettidinescuetal2014,is2015}.

The cluster deserves more of our attention in order to unveil its actual 
origin, particularly in the light of  the implications that such an unusual 
object could have for our understanding of the cluster formation in the LMC. 
LMC outer disc clusters have long been commonly thought to be old 
and hence, to be key to reconstructing the early galaxy formation and chemical 
enrichment history \citep{piattietal2009,getal10}. However, the recently 
discovered cluster does not comply with such a picture, because it seems to be younger 
(age $\sim$ 280 Myr, Piatti 2016) than what should be expected. Likewise, the outer disc is 
commonly featured as a more metal-poor structure ([Fe/H] $\la$ -0.5 dex) than 
the inner LMC body \citep{hz09,meschin14}, although the new cluster 
([Fe/H] $\sim$ -0.1 dex, Piatti 2016) is at the metal-rich end of the LMC cluster 
metallicity distribution, making it a really odd object.

A network of streams surrounding the LMC has been discovered
\citep[see, e.g.][]{bk2016,mackeyetal2016} that could be relics of collisions between 
both MCs \citep{hammeretal2015,salemetal2015} and possibly the MW. 
However, no cluster 
has been discovered so far in those streams \citep[e.g.,][]{mbetal2017}, except for 
those in the Magellanic Bridge \citep[e.g.,][]{bicaetal2015,petal15a}, which is located
on the opposite side of the LMC with respect to the newly discovered cluster. 
Moreover,
the new cluster is located towards a direction where no stream has been seen yet.
Although the spatial distribution of the latest discovered clusters demonstrates 
that the outer regions of the MCs were less explored in the past 
\citep[see, e.g.,][]{siteketal2016,pieresetal2016,p17a}, the aforementioned newly 
discovered cluster is 
½positioned at an LMC-centric distance.
positioned at a LMC-centric distance
where no clusters are expected to have formed.

To unveil the origin of this potentially very unusual cluster, 
we have conducted a photometric and spectrocopic study from which we
comprehensively derive its astrophysical properties. The observations are
described in Section 2, while radial velocity (RV) and metallicity
membership probabilities are presented in Section 3. In Section 4 we estimate
the cluster's astrophysical properties and discuss their implications for the
cluster origin. We found that the cluster is not as extreme as
originally estimated by \citet{p16}. Finally, the main conclusions of this work are summarized 
in Section 5.

\section{Observational data}

\subsection{SOAR/SAMI observations}

The cluster was observed on January 3, 2018 with the SOAR Adaptive Module (SAM) 
coupled with its Imager (SAMI), installed at the SOAR 4.1m telescope at 
Cerro Pach\'on, Chile. SAM is a ground-layer adaptive optics system correcting 
atmospheric turbulence near the ground. Technical details of the instrument 
can be found in \citet{tokovinin16}. SAMI provides a field-of-view (FOV) of 
$3\times3$ arcmin$^2$ with a pixel scale of 0.091\arcsec when used with 
a 2$\times$2 binning. Observations were taken in the SDSS $g$ and $i$ filters, 
with 3$\times$300 and 3$\times$200 sec exposures, at
mean corrected FWHMs of the stellar images of 0.54\arcsec\, in $g$ and
 0.41\arcsec\, in $i$, respectively.

\begin{figure*}
   \includegraphics[width=\columnwidth]{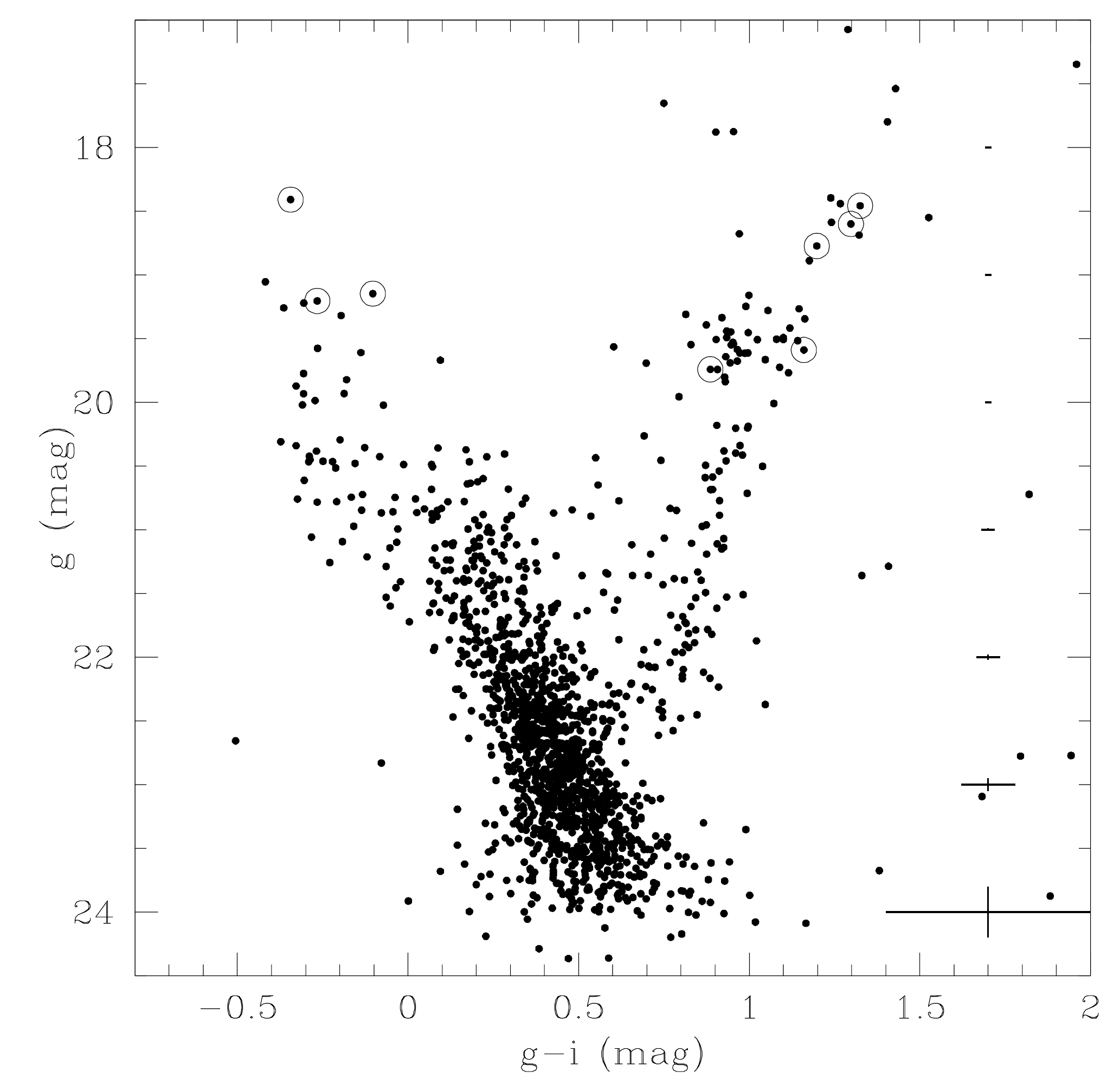}
   \includegraphics[width=\columnwidth]{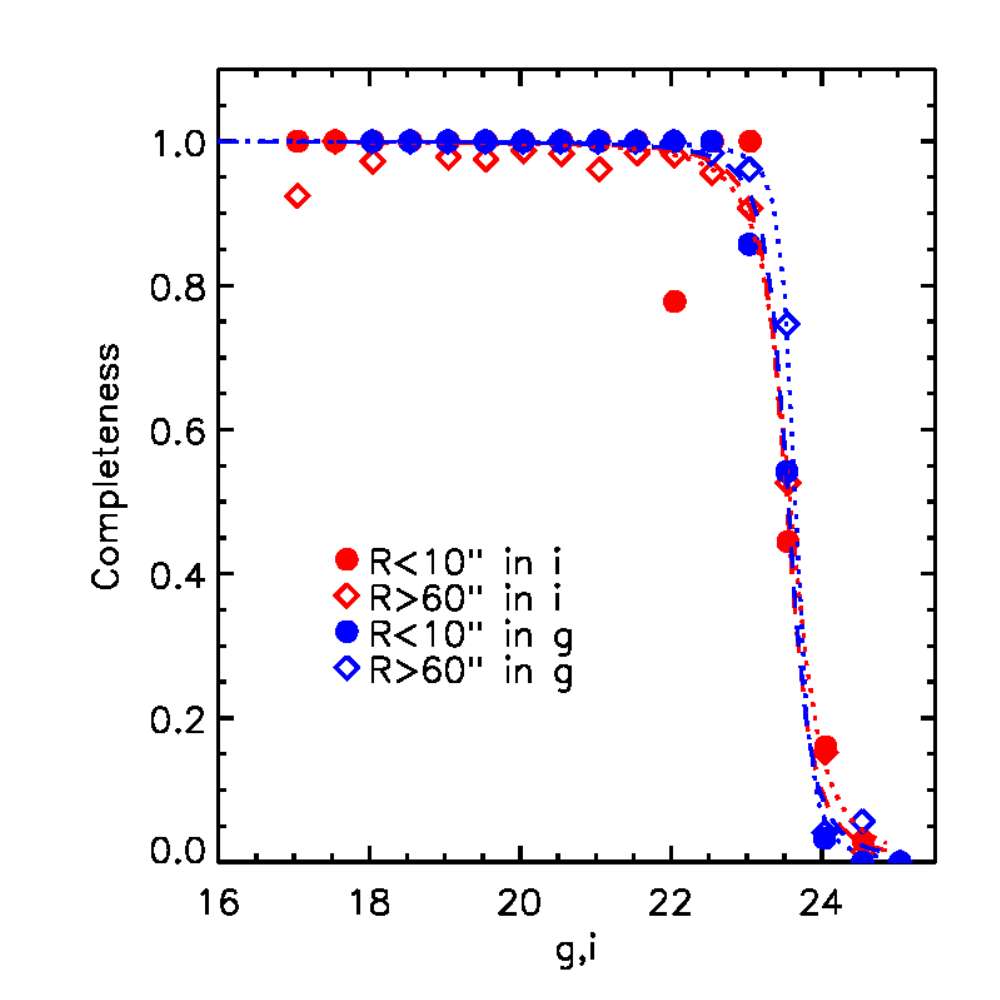}
    \caption{Observed CMD of stars in the field of the new star cluster (left panel) and
the completeness curves obtained from our photometry (right). Photometry errors
are represented by error bars at the CMD's right margin, and spectroscopic targets are encircled.}
   \label{fig:fig1}
\end{figure*}

\begin{table*}
\caption{$gi$ data of stars in the field of the new star cluster$^a$.}
\label{tab:table1}
\begin{tabular}{@{}lcccccccccc}\hline
Star & R.A. & DEC. & $g$ & $\sigma(g)$ & $i$ & $\sigma(i)$ & $\chi(g)$ & sharp($g$) & $\chi(i)$ & sharp($i$) \\
     & (deg) & (deg) & (mag) & (mag) & (mag) & (mag) & & & &  \\\hline
 -- & --& --& -- & --& --& -- & --& --& -- & --  \\
   626& 90.89030115& -72.42108798&  23.510&  0.087&  23.046&  0.055&  1.006& -0.304&  1.079&  0.400\\
   627& 90.88438994& -72.42100738&  22.000&  0.022&  21.521&  0.015&  0.986&  0.062&  1.061& -0.194\\
   628& 90.94368341& -72.42106770&  22.168&  0.026&  21.364&  0.013&  1.015&  0.225&  1.013&  0.165\\
 -- & --& --& -- & --& --& -- & --& --& -- & -- \\
\hline
\end{tabular}

\noindent $^a$A portion of the table is presented here for guidance of its
contents. The entire table is available as Supplementary material
in the online version of the journal.
\end{table*}

\subsubsection{Data reduction and photometry}

SAMI images were reduced employing the multi-instrument imaging pipeline
 \textsc{theli}\footnote{$^1$ https://www.astro.uni-bonn.de/theli/gui/background.html}
 \citep{schirmer13}, by applying bias subtraction and flat-fielding the images with bias and 
twilight sky frames taken during the same
 night. After flat-fielding, a difference of a few percent was still visible
 in the background between the four SAMI amplifiers. This was removed 
applying a collapse correction implemented within 
\textsc{theli}.

The images were not combined. Instead the photometry was done in each of 
them separately with the {\sc daophot/allstar} suite of programs \citep{stetson87}. 
The procedure involved the following steps: a) adding back to each image a 
mean sky background level that had been subtracted during the collapse correction.
 This is necessary for {\sc daophot} to have the proper photon statistics. b) Sources 
in each field were searched with SExtractor \citep{bertin96} combining a 
Gaussian filter and a ``Mexican hat'' filter; an approach that enhances the 
detection of both faint sources and sources near very bright ones 
\citep[e.g.][]{salinas15}. c) Nearly 40 bright and reasonably isolated
 point sources were selected in the image with the best image quality,
 in order to model the point-spread-function (PSF) as a quadratically varying  Moffat function as performed 
in previous studies with the same instrument \citep{fraga13,salinas16}. 
The resulting PSF subtracted images revealed no significant increase of the residuals 
as function of radius across the whole small FOV ($3\times3$ arcmin$^2$).
Finally,
the psf photometry was obtained using the {\sc allstar} task \citep{stetson87}, and all 
measurements per filter were averaged using {\sc daomaster}'s routines \citep{stetson92}.
 Aperture corrections were calculated based on a subset of psf stars
and instrumental magnitudes converted into standard ones using 
transformation equations derived from observations of standard star fields 
observed during the same night. The transformation equations were fitted with
rms of 0.011 mag and 0.012 mag for $g$ and $i$, respectively, as given by {\sc iraf/photcal}. 
Table~\ref{tab:table1} shows a portion of the
resulting photometric data set,
while Fig.~\ref{fig:fig1} depicts the colour-magnitude diagram (CMD) built from
all the measured stars.

\subsubsection{Photometric completeness}

The photometric completeness was derived from standard artificial star experiments on all observed
images using the {\sc daophot.addstar} task. To do this, we added 300 stars with $i$ magnitudes between 16 and 25 mag, following an exponential distribution in order to achieve a better statistics at the
fainter limit. Two thirds of the stars were randomly distributed within the SAMI FOV, while the remaining
one third was distributed following a two dimensional Gaussian distribution centered on the cluster. 
The initial $g$ magnitudes for the {\sc addstar} experiments were calculated
from the $i$ magnitudes using a polynomial fit to the main features in the CMD. The experiment
 was repeated 30 times per observed image for a total of 
9000 artificial stars. All frames with the added stars were photometered exactly as the original
 frames. Input and output catalogues were matched using a combination of {\sc daomaster} \citep{stetson92} 
and STILTS \citep{taylor06}.

Fig.~\ref{fig:fig1} shows the results of the completeness experiments for a radial bin with $R<10\arcsec$ and $R>60\arcsec$. Differences in radial completeness are almost negligible and a very small difference is seen between the filters. Additionally, the results were fitted with a "Pritchett function" 
\citep{mclaughlin94}, commonly used in globular cluster systems studies \citep[e.g.][]{salinas15}, which can be seen as dotted and dash-dotted lines, respectively.

\subsection{GMOS-S/MOS observations}

Spectroscopic observations were obtained with the Gemini Multi-Object 
Spectrograph \citep[GMOS,][]{hook04}, mounted on the 8.1m Gemini South 
telescope, located on Cerro Pach\'on, Chile, on the nights of April 20 and 21, 2018.
 GMOS is equipped with three Hamamatsu CCDs \citep{gimeno16} yielding a total FOV of 
5.5$\times$5.5 arcmin$^2$. Given the very compact nature of the new star cluster, two 
masks were designed, including 5 and 3 stars, respectively, selected from 
\citet[][see some examples in figure 2]{p16}. 
Observations were taken with the R831 grating, centered at 850 nm, providing 
a resolving power of $R\sim2200$ for 1\arcsec\, slitlets. Exposures of 3$\times$900 sec were
 obtained for each mask.

\begin{figure}
   \includegraphics[width=\columnwidth]{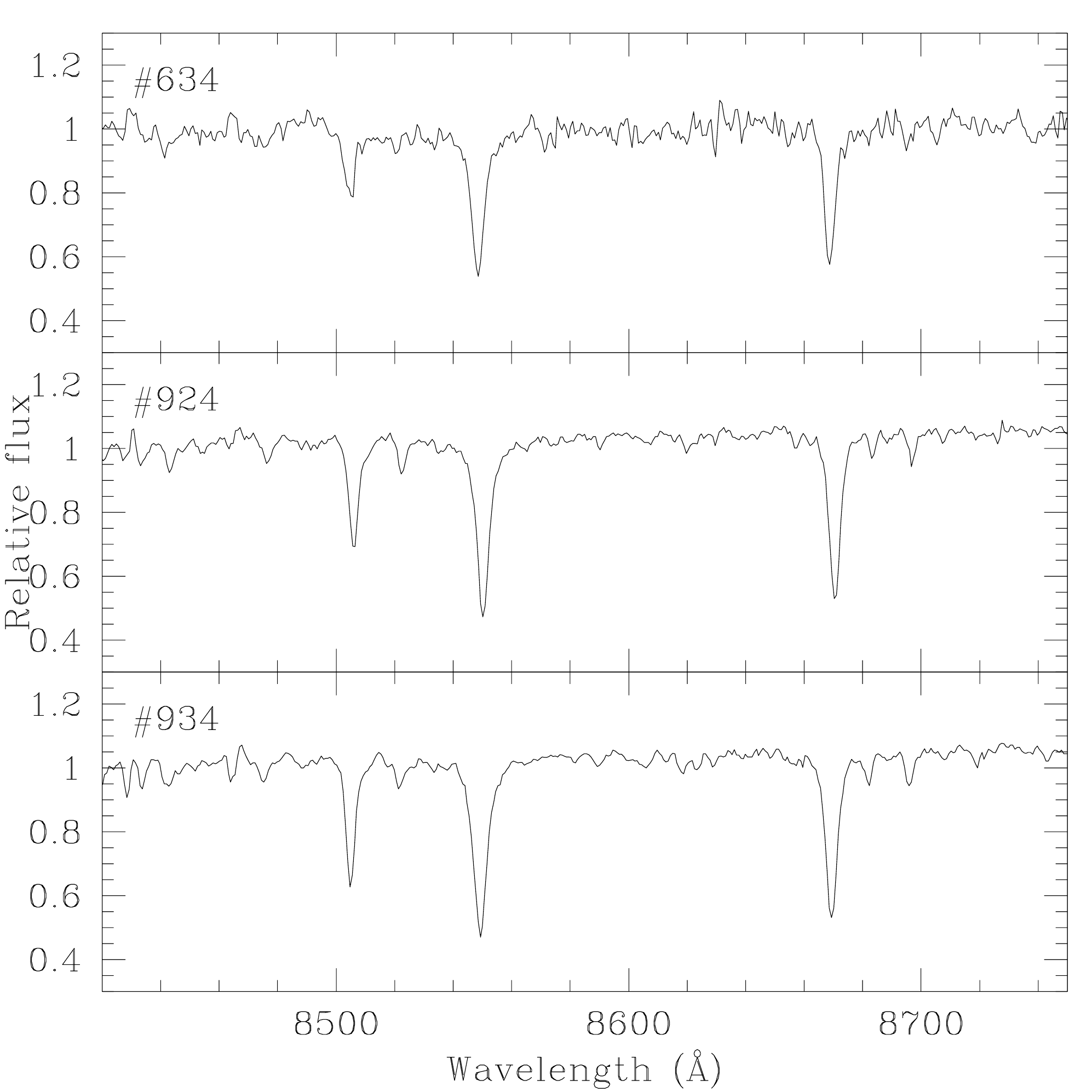}
    \caption{Normalised spectra of some observed stars. Star numbers are as in
Table~\ref{tab:table2}.}
   \label{fig:fig2}
\end{figure}

Data reduction was done with standard techniques implemented within the Gemini 
IRAF\footnote{IRAF is distributed by the National 
Optical Astronomy Observatories, which is operated by the Association of 
Universities for Research in Astronomy, Inc., under contract with the National 
Science Foundation.} package. After bias subtraction, flat-fielding and wavelength calibration
 with CuAr spectra taken right after the science observations, the zero point 
of the latter was corrected by measuring the position of bright, isolated 
skylines near the CaT region identified from a UVES high-resolution spectrum
 \citep{hanuschik03}. The shifts in the zero point were found to be no larger 
than 0.5\AA. After this correction, spectra were sky-subtracted, extracted and
 combined following standard methods. Fig.~\ref{fig:fig2} illustrates some of the
obtained spectra.

\section{Radial velocity and metallicity measurements}

RVs were measured with the cross-correlation algorithm of \citet{tonry79} 
as implemented in IRAF {\sc fxcor} task. Forty spectra were selected from the 
library of \citet{cenarro01} from a list of 706 stars observed in the 
8348$-$9020\AA\,region with a resolution of 1.5\AA. This template subsample 
included stars with late O to late M spectral types. Each science spectrum was
then cross-correlated against the whole template subsample, and the five 
measurements with the highest cross-correlation peak were averaged to obtain 
a final RV per star, which include heliocentric corrections. 
Table~\ref{tab:table2} shows the measured signal-to-noise (S/N) ratio in the
region of the CaII triplet lines and the resulting RVs with their uncertainties.

\begin{table*}
\caption{Radial velocities, CaII triplet line equivalent widths and metallicities of selected stars.
Spectra with S/N = 15 were only useful to measure RVs.}
\label{tab:table2}
\begin{tabular}{@{}lcccccc}\hline
Star$^a$ &  S/N & RV  &  W8498  &  W8542  &   W8662  & [Fe/H] \\\hline                 
634& 32 &216.9$\pm$2.0  &   1.131$\pm$0.012&  3.552$\pm$0.128 & 2.177$\pm$0.029 & -0.48$\pm$0.20\\
724& 45 &275.9$\pm$2.1  &   1.208$\pm$0.053&  3.504$\pm$0.118 & 2.221$\pm$0.106 & -0.52$\pm$0.23\\
748& 60 &218.0$\pm$2.0  &   1.450$\pm$0.050&  3.623$\pm$0.045 & 2.776$\pm$0.031 & -0.46$\pm$0.17\\
796& 15 &259.4$\pm$6.6  &   --- & --- & --- & --- \\
836& 15 &211.0$\pm$7.0  &   --- & --- & --- & --- \\
868& 15 &305.0$\pm$11.0 &    --- & --- & --- & --- \\
924& 65 &275.3$\pm$2.0  &   1.838$\pm$0.095&  3.907$\pm$0.033 & 2.736$\pm$0.040 & -0.18$\pm$0.19\\
934& 60 &238.8$\pm$2.0  &   1.909$\pm$0.134&  3.801$\pm$0.059 & 2.692$\pm$0.040 & -0.30$\pm$0.21\\\hline
\end{tabular}

\flushleft $^a$ Star numbers are as in Table~\ref{tab:table1}.
\end{table*}

\begin{figure}
   \includegraphics[width=\columnwidth]{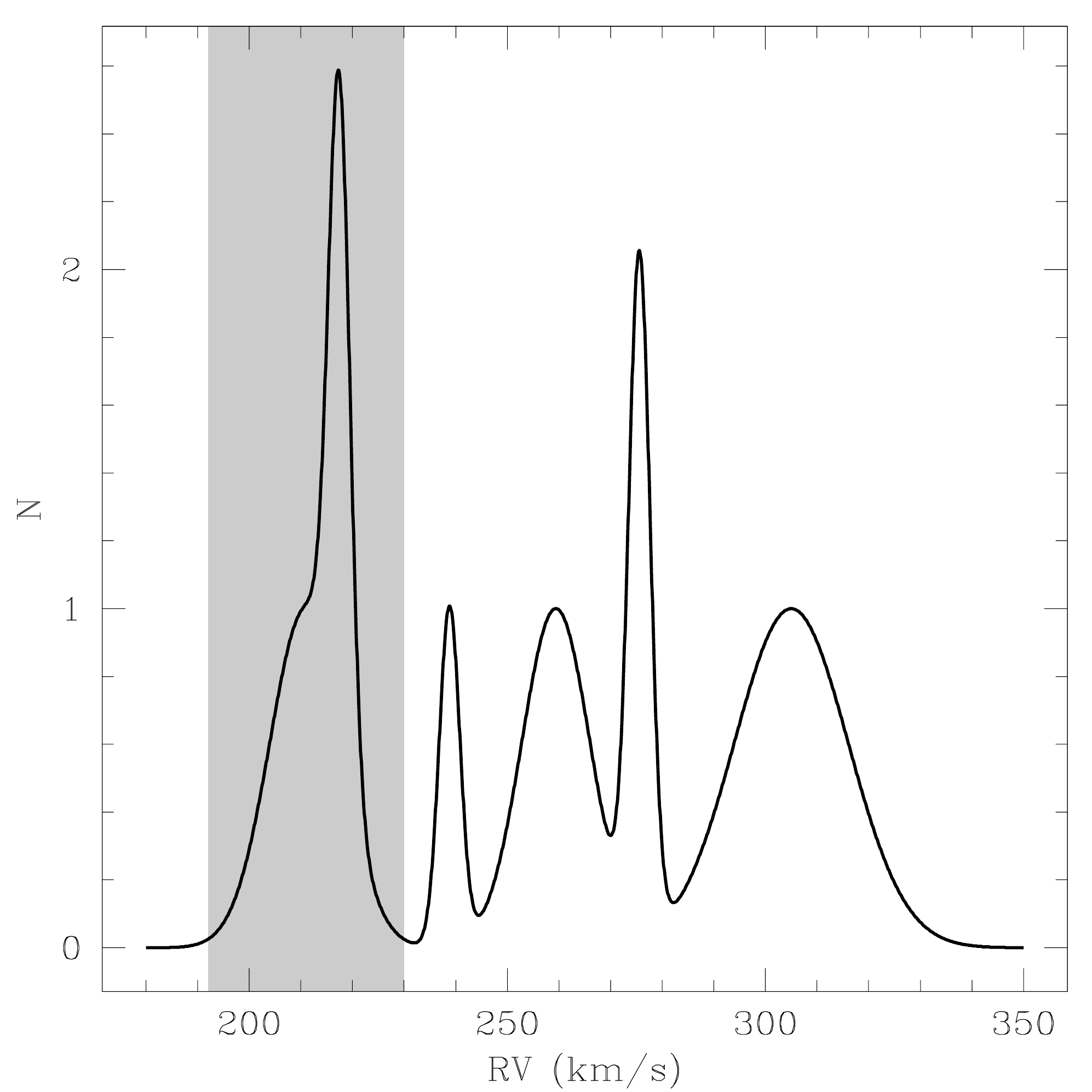}
    \caption{RV distribution function for stars observed in the field of
the new star cluster. The shaded region represents the adopted cluster
RV range.}
   \label{fig:fig3}
\end{figure}

The RV distribution function was then constructed by summing all the individual RV values.
We represented each RV value by a Gaussian function with centre and full-width half 
maximum equal to the mean RV value and  2.355 times the associated error, respectively (see 
Table~\ref{tab:table2}), and assigned to each Gaussian the same amplitude. 
Fig.~\ref{fig:fig3} shows the resulting RV distribution function, where the shaded region 
represents the cluster RV range adopted on the basis of the largest number of stars with 
similar RV measurements. Hence, the mean cluster RV computed from stars \#634, 748 and 836 
turned out to be 215.3$\pm$3.1 km s$^{\rm -1}$.

\begin{figure}
   \includegraphics[width=\columnwidth]{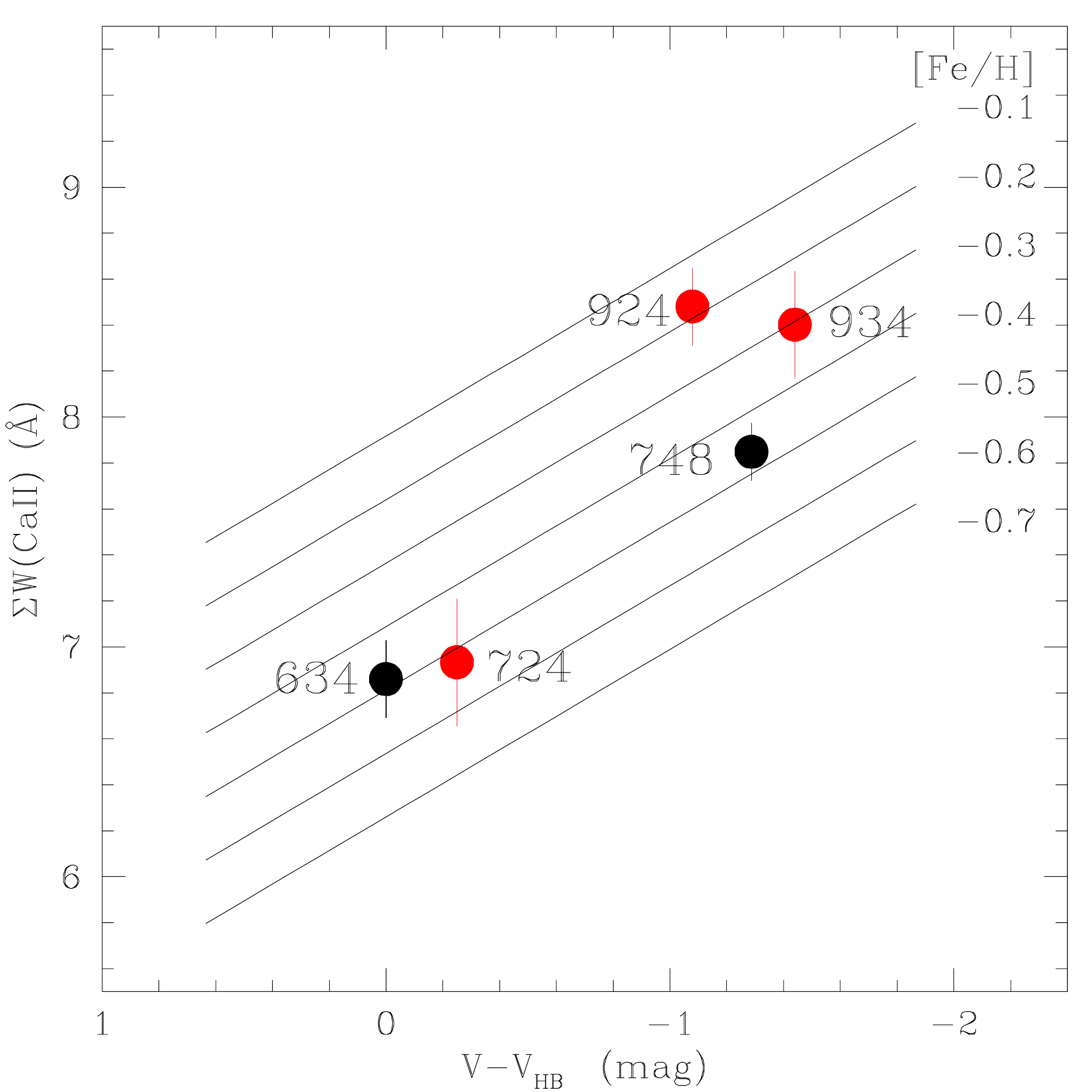}
    \caption{Sum of the CaII triplet line equivalent widths as a function of
$V-V_{HB}$ for stars observed in the field of the new star cluster. The star numbers
according to Table~\ref{tab:table2} and the error bars are included. Black
and red filled cirles represent probable cluster member and non-member stars,
respectively (see text for details).
Iso-abundance lines derived by \citet{coleetal2004} for some [Fe/H] values 
are also depicted. }
   \label{fig:fig4}
\end{figure}

We then measured equivalent widths of the CaII infrared triplet lines of the
observed red giants using the {\sc splot} 
package within IRAF. Their resulting values and uncertainties are listed in 
Table~\ref{tab:table2}. The latter were estimated by considering upper and lower envelopes
of the continuum at both sides of the CaII lines. \citet{piattietal2018} showed
that this procedure of measuring CaII lines equivalent widths is highly consistent with no
systematic offset with respect to the techniques employed by \citet{coleetal2004,coleetal2005,dacosta2016},
among others.
Then we overplotted the sum
of the equivalent widths of the three CaII lines ($\Sigma$W(CaII)) in the 
$\Sigma$W(CaII) versus $V-V_{\rm RC}$ plane, which has been calibrated in terms of 
metallicity  \citep[see, e.g.,][]{coleetal2004,coleetal2005}. In that diagram $V_{\rm RC}$ refers
to the mean magnitude of the cluster red clump. In this case, we first
converted $g$ magnitudes into Johnson $V$ ones using the expression:

\begin{equation}
V = g - 0.3557*(g-i) - 0.0600,
\end{equation} 

\noindent which we obtained from Lupton's 
(2005)\footnote{http://www.sdss.org/dr13/algorithms/sdssUBVRITransform/\#Lupton2005.} 
transformation equations between $ugriz$ and $UBVR_cI_c$ photometric systems and the relationship 
$g-i = (g-r) + (r-i)$. We adopted for the new star cluster $g_{\rm RC}$ = 19.74 mag 
(see  Fig.~\ref{fig:fig6}). Fig.~\ref{fig:fig4} shows the resulting $\Sigma$W(CaII) versus 
$V-V_{\rm RC}$ diagram, where we included iso-abundance lines according to
eq. (5) of \citet{coleetal2004} for $\beta = 0.73$\,\AA/mag, while the interpolated [Fe/H] 
values are listed in the last column of Table~\ref{tab:table2}. Note that
the influence of age on eq. (5) is relatively small, so that it can be applied to stellar systems with 
ages as young as $\sim$ 0.3 Gyr 
\noindent \citep[see][]{coleetal2004,carreraetal2007,dacosta2016}. 
We calculated the [Fe/H] 
errors by propagating those of the coefficients in eq. (5) and  $\sigma$($\beta$)
\citep{coleetal2004}, the photometric errors (Table~\ref{tab:table1}) and 
$\sigma(\Sigma$W(CaII)) (Table~\ref{tab:table2}), respectively. As can be seen, 
the likely red giant RV members (stars \#634 and 748) also share a very similar metallicity 
of  $\sim$ -0.5 dex. The remaining red giants have interpolated [Fe/H] values 
falling well within the LMC field star metallicity range \citep{coleetal2005,pg13}, 
but cover a metallicity range too wide to be consistent with cluster membership, further
strenghtening our conclusions regarding the cluster parameters.

\section{Analysis and discussion}

The mean star cluster metallicity helped us to select the appropriate theoretical
isochrone to match the cluster CMD. In doing that, we used the well-established
colour excess $E(B-V)=0.08\pm$0.01 mag - also confirmed from the NASA/IPAC 
Extragalactic Data base\footnote{http://ned.ipac.caltech.edu/. NED is operated by the 
Jet Propulsion Laboratory, California Institute of Technology, under contract with 
NASA.} (NED) -,  as well as one upper main sequence (MS) and  the two red giant RV members 
to better constrain the cluster age and distance. 

\begin{figure*}
   \includegraphics[width=\columnwidth]{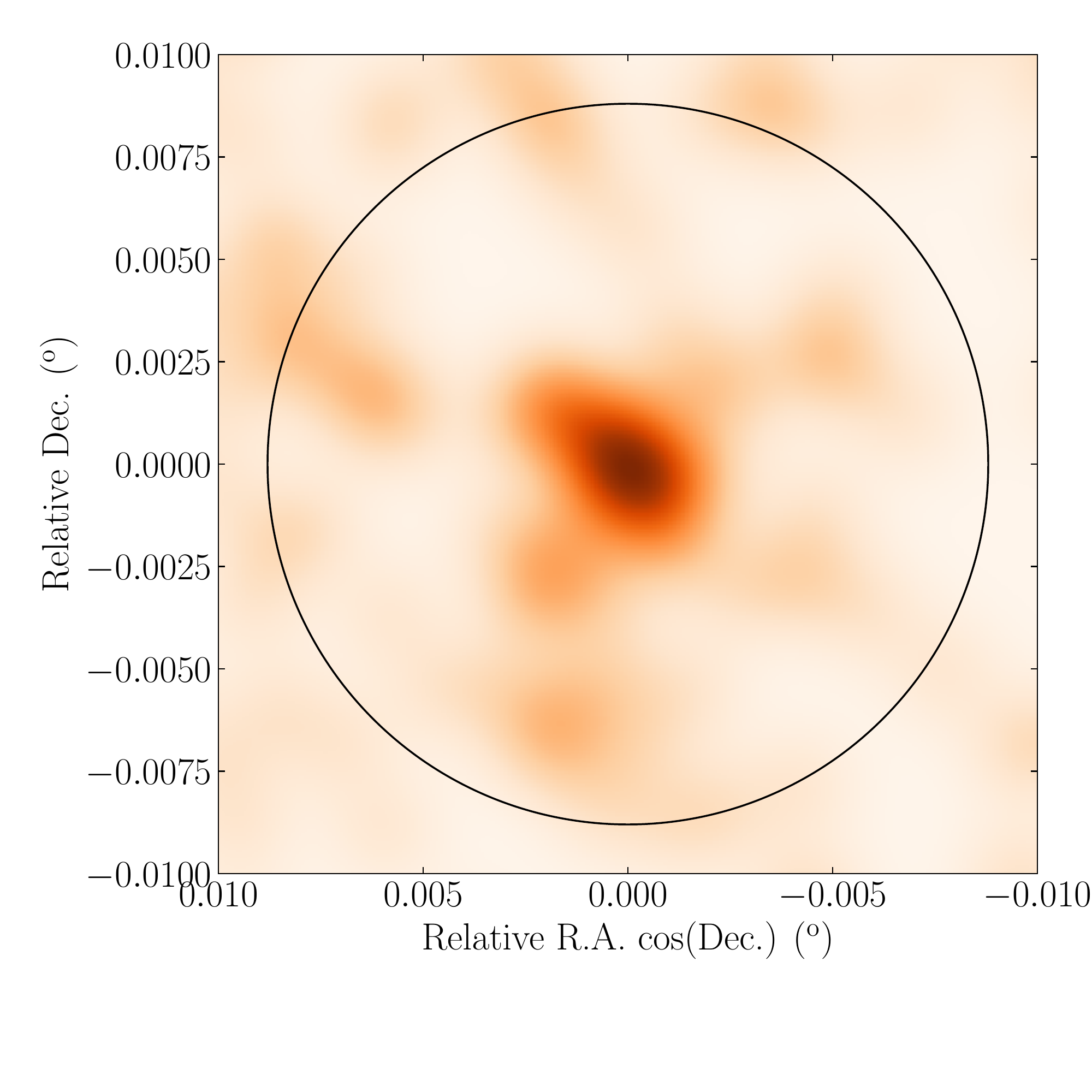}
   \includegraphics[width=\columnwidth]{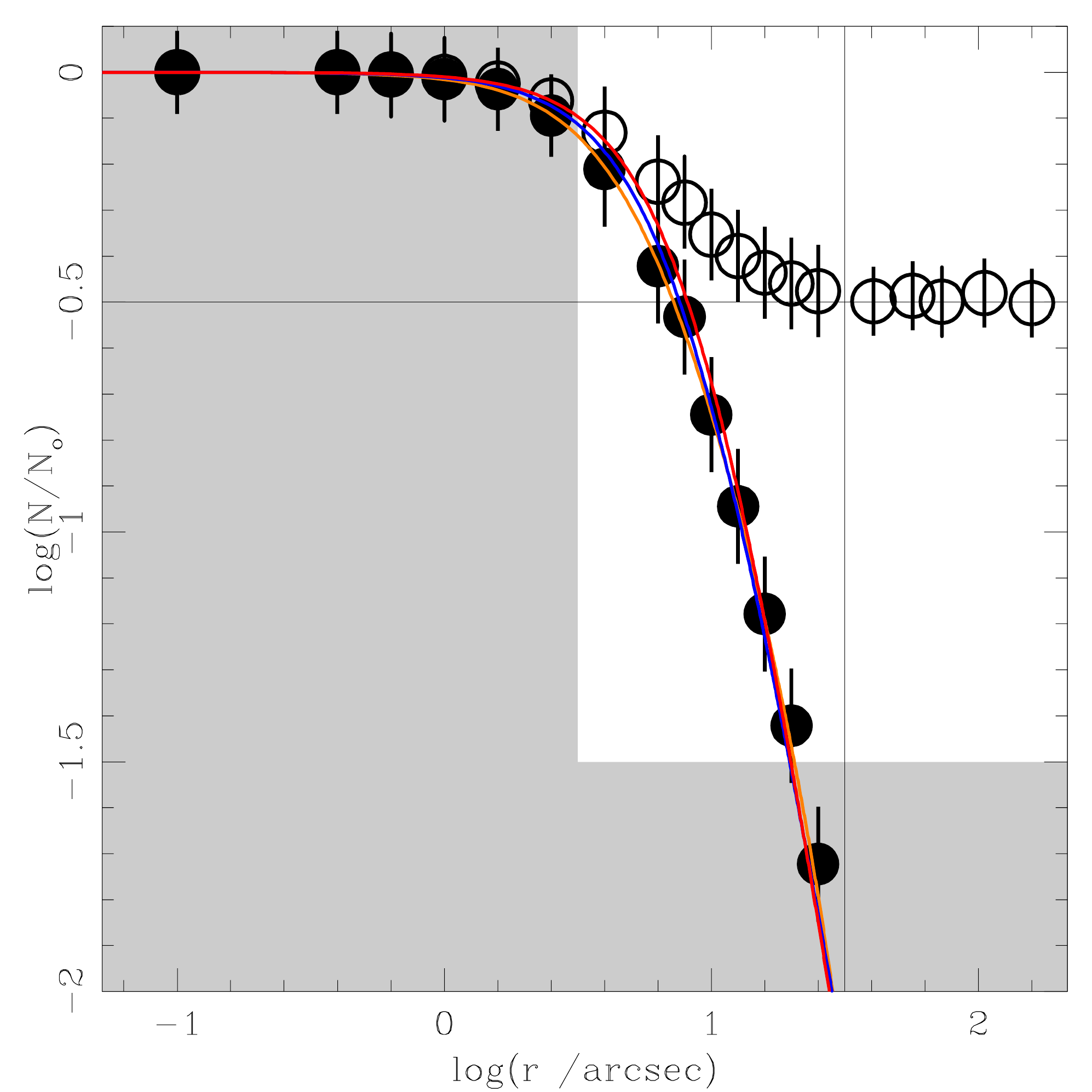}
    \caption{{\it Left: }Stellar density map centred on the new star cluster,
smoothed by a Gaussian kernel with standard deviation 3.6 arcsec. The black circle
represents the cluster radius. {\it Right :}
Normalised observed and background-subtracted cluster stellar radial profiles
depicted with open and filled circles, respectively. Error bars are also included.
The horizontal and vertical lines represent the mean background level and the adopted
cluster radius, respectively.
The best-fitted \citet{king62}, \citet{eff87} and \citet{plummer11} models are superimposed
with solid orange, blue and red lines, respectively. The shaded area represents the region not
considered by \citet{p16}.}
   \label{fig:fig5}
\end{figure*}

With the aim of reproducing the cluster CMD, we first built the cluster stellar radial 
profile from which we derived its radius, namely: the intersection of the cluster
density profile with the mean background level.
For distances smaller than 10 arcsec from the cluster centre, the cluster radial
profile was built by averaging the generated values in the stellar density map constructed 
from a kernel density estimator (KDE) technique. As can be seen in Fig.~\ref{fig:fig5},
the stellar density map resulted in a nearly smoothly continuing distribution of  
density values that allowed us to trace more accurately the cluster radial profile in
the innermost regions, instead of using small boxes to count stars inside them. 
Star counts within appropriate size boxes were more
suitable to delineate the outer regions of the cluster profile as well as the
background level, because of the lack of spurious fluctuations due to the small 
number statistics. 

We run both the KDE routine and the counting of stars over a subsample of stars with $g$ 
magnitudes brighter than those for the 90 per cent completeness level (see 
Fig.~\ref{fig:fig1}). The KDE was repeatedly run by employing a Gaussian function with 
bandwidths between 1.8 and 4.8 arcsec in steps of 0.6, while the star counts were performed 
by using boxes with sizes between 4.0 and 8.0 arcsec in steps of 1.0 arcsec, respectively. 
The resulting averaged cluster radial profile is shown in Fig.~\ref{fig:fig5} (open circles)
with the respective  error bars. The mean background level was then
estimated by averaging those values for log($r$ /arcsec) $\ge$ 1.7 (horizontal line in 
Fig.~\ref{fig:fig5}). We chose this limit because it is readily visible from the figure
that a meaningful mean background can be obtained from the values beyond it.
Hence, the intersection of cluster radial profile with that mean background level resulted 
to be at log($r$ /arcsec) = 1.55$\pm$0.10 (31.6$^{+8.0}_{-6.0}$ arcsec), in excellent agreement
with the cluster radius estimated by \citet{p16}. Note that all the stars observed
spectroscopically  lie inside the cluster radius.

The background-subtracted cluster radial profile (filled circles in Fig.~\ref{fig:fig5}),
with error bars that come from considering in quadrature the uncertainties of the measured 
profile and the dispersion of the background level, 
surpasses that previously traced by \citet{p16}, in the sense that it reaches the innermost 
regions of the cluster. This is mainly because the present cluster profile was built
from a nearly continuous density map rather than from star counts. Likewise, we now reach 
fainter outermost cluster structures, despite the fact that we constrained the sample of 
stars to those with $g \la 22.0$ mag, for completeness reasons. For the sake of the reader, 
we have shaded these regions in the figure. By using the present more extended 
cluster radial profile, we followed the procedure described in detail by \citet{pm2018} to 
fit \citet{king62}, \citet{eff87} and \citet{plummer11} models and thus to obtain accurate
cluster structural parameters. We derived a cluster core radius ($r_c$) of 5.5$\pm$0.5 arcsec, 
a  half-light radius ($r_h$) of 12.0$\pm$1.0 arcsec, a tidal radius ($r_t$) of 55.0$\pm$5.0 
arcsec and a  power-law index ($\gamma$, the slope at large radii in Fig.~\ref{fig:fig5}) of 
3.5$\pm$0.2, respectively.

\begin{figure}
   \includegraphics[width=\columnwidth]{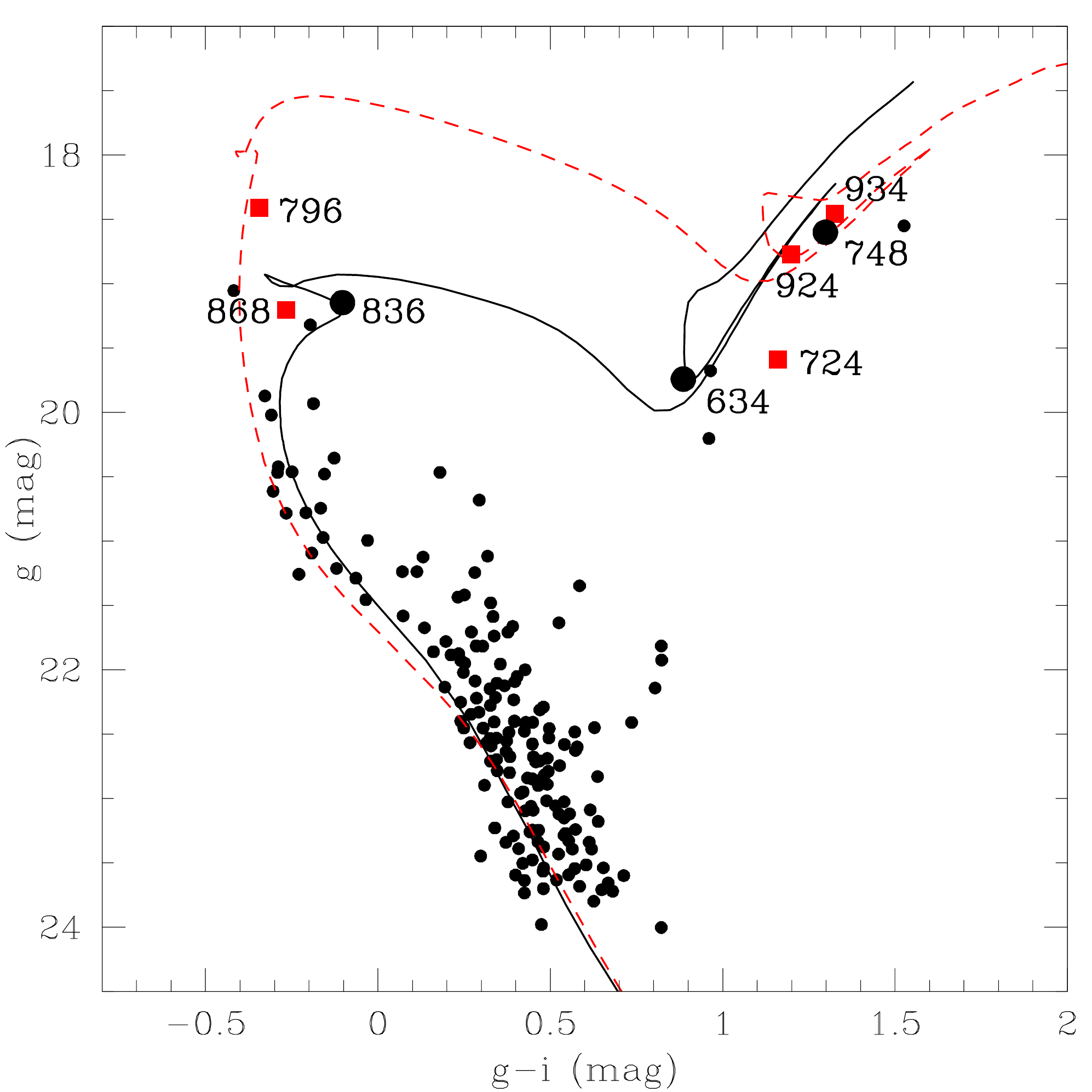}
    \caption{CMD of the observed stars located within the cluster radius. Red boxes and big black 
filled circles represent non-members and members according to their RVs and metallicities;
their star numbers are also indicated (see Table~\ref{tab:table2}). Theoretical isochrones
of \citet{betal12} for the the present best solution (log($t$ /yr)=8.95, [Fe/H]=-0.4 dex)
and that of \citet{p16}  (log($t$ /yr)=8.45, [Fe/H]=-0.1 dex) are superimposed
with black  solid and red dashed lines, respectively.}
   \label{fig:fig6}
\end{figure}

Fig.~\ref{fig:fig6} shows the extracted cluster CMD that results from using all the stars
measured within the cluster radius. We have not carried out any star field decontamination
so that, besides the cluster upper MS and red clump, the composite LMC star field MS at
fainter magnitudes and few giant stars are also superimposed. We have drawn with bigger
symbols stars for which we measured RVs, distinguishing those members from
non-members with black filled circles and red boxes, respectively. At first glance, 
the three cluster members are fortunately placed in key positions as to perform
a sound fitting of theoretical isochrones. To do that, we used as entries the known cluster
reddening mentioned above and the spectroscopic metallicity derived in Section 3,
allowing small variations according to their uncertainties; the only free parameters being 
the cluster age and distance modulus. As for the theoretical isochrones we used those computed by 
\citet{betal12}. 

From a grid of values of age and distance modulus spread in the same ranges as in \citet{p16}, 
we found that the
isochrone which best resembles the cluster features, namely: the upper MS, the MS turnoff from 
star \#836, the base of the red clump from star \#634 and the red giant branch from star \#748,
is that of log($t$ /yr) = 8.95$\pm$0.05 (0.89$^{+0.11}_{-0.10}$ Gyr) and $m-M_o$ = 
18.40$\pm$0.05 mag (d= 47.9$\pm$1.1 kpc), for a metallicity of [Fe/H] $= -0.4\pm$0.1 dex. 
The resulting cluster
properties reveal that the object is not as young, metal-rich or located as far away
from the LMC
as inferred by \citet{p16}, namely, age= 0.28$^{+0.04}_{-0.03}$ Gyr, [Fe/H]= -0.10$\pm$0.05 dex
and d=60.3$\pm$1.4 kpc, but one that lies in the main body of the galaxy with an age and 
a metal content that agree very well with the known picture of the LMC cluster
age-metallicity relationship \citep{pg13}.  The inferred metallicity is in agreement with
the [Fe/H] values and uncertainties of stars \# 634 and 748 (see Table~\ref{tab:table2}). 
Fig.~\ref{fig:fig6} shows with a black solid line
the best solution of the isochrone fitting, while the red dashed line represents the
solution found by \citet{p16}. From the figure, it becomes clearer that the previous 
younger age, and hence larger distance and metallicity, come from considering a much brighter
cluster MS turnoff and red clump. Here, the RV measurements allowed us to resolve
the conundrum of its previously derived astrophysical properties. 

\begin{figure}
   \includegraphics[width=\columnwidth]{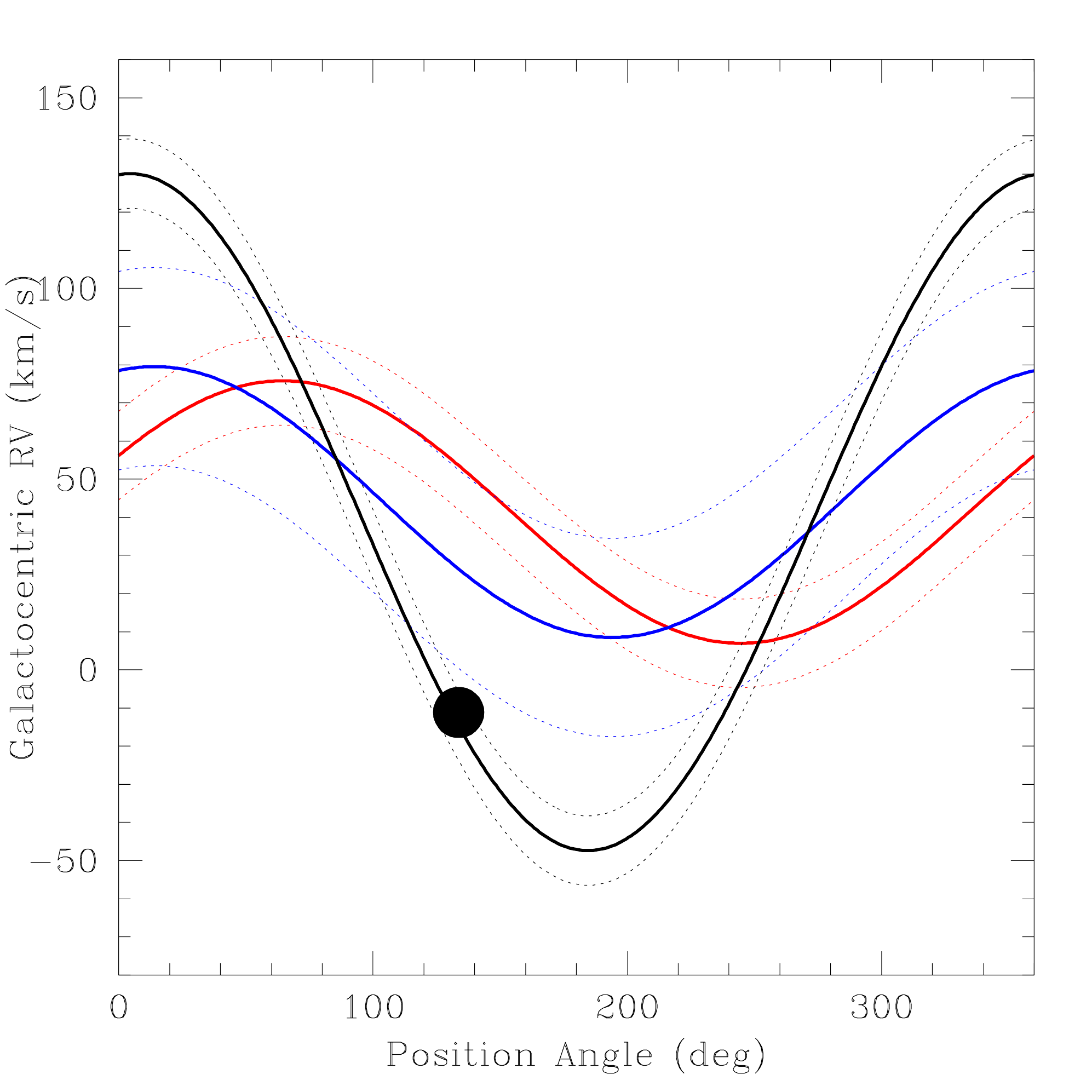}
    \caption{Position of the new star cluster in the Galactocentric RV versus PA diagram 
(filled circles).  Error bars are smaller that the symbol size. The best solution for a disc 
that contains it is drawn with a black solid line, while the black dotted lines represent 
those considering every involved errors. We included the curves derived 
by \citet[][solution \#3 in their Table~3]{s92} from mostly outer LMC clusters and by 
\citet{vdmk14} from $HST$ proper motions of 22 LMC fields and 723 young field stars 
with blue and red solid lines, respectively, for comparison purposes. We include 
the derived errors in the rotational curves with dotted lines.}
   \label{fig:fig7}
\end{figure}

We used the mean star cluster RV and its position angle (PA) to complementary assess
the possible association of the cluster with the LMC disc 
\citep[see][]{s92,getal06,shetal10,vdmareletal2002,vdmk14}. The cluster PA and its
deprojected distance ($\rho$) were calculated considering a disc 
centred at R.A.$=80.05\pm0.34 \degr$ and Dec. $=-69.30\pm0.12 \degr$,  with an
inclination $i=26.2\pm5.9 \degr$ and PA of the line-of-nodes $\Theta=154.5\pm2.1 \degr$, 
obtained by \citet{vdmk14} from $HST$ average proper motion 
measurements for stars in 22 fields and RVs of 723 young LMC field stars. 
We derived PA$=131.4\pm3.0 \degr$ and $\rho= 4.63\pm0.03 \degr$, which implies 
that the new star cluster is located in the outer disc \citep{betal98}. We then converted the mean 
heliocentric cluster RV to the Galactocentric one through eq.(4) in \citet{fw79}, in order
to use the disc solutions found by \citet{s92}. Fig.~\ref{fig:fig7}
shows the position of the new star cluster in the RV versus PA plane (filled circle)
with the mean rotation curve obtained by \citet[][solution \#3 in their Table~3]{s92} using 
mostly outer LMC star clusters drawn with a blue solid line. In the figure, dotted blue
lines represent the above solution with the uncertainties in the circular and systemic 
velocities and the derived dispersion velocity. The new star cluster
appears to rotate in a slightly different disc than that containing most of the outer
LMC star clusters. Likewise, it rotates even more differently than the young stellar
population, as judged by the position and shape of the mean disc solution and those with 
its uncertainties and velocity dispersion found by \citet{vdmk14} mentioned above, which we 
drew with red solid and dotted lines, respectively.

In order to characterise the disc that fully contains the new star cluster, 
 we looked for the circular velocity ($v_{\rm rot}$) and PA of the 
line-of-nodes (PA$_{\rm LOS}$) of a disc that rotates at the deprojected cluster distance with
 the corresponding cluster RV.
We evaluated eq.(1) of \citet{s92} at the cluster PA for different values of
($v_{\rm rot}$,PA$_{\rm LOS}$) and then minimized by $\chi^2$ the difference between the
calculated and measured cluster RV. We used a grid of $v_{\rm rot}$ from 0.0 up to 200.0 km s$^{\rm -1}$ 
in steps of 1.0 km s$^{\rm -1}$, and a range of PA$_{\rm LOS}$ from 0.0 up to 180.0 degrees in steps of 
1.0 degree, besides the uncertainties in the cluster RV and PA. Thus, the most suitable disc 
turned out to be that with $v_{\rm rot}=177.5\pm11.4$ km s$^{\rm -1}$ and PA$_{\rm LOS}=
94.5\pm 2.7 \degr$.
This solution with its dispersion is drawn in Fig.~\ref{fig:fig7} with black solid
and dotted lines, respectively.  When comparing our results with those for the outer
LMC star clusters found by \citet{s92}, i.e, $v_{\rm rot}$=71$\pm$8 km s$^{\rm -1}$ and
PA$_{\rm LOS}$ = 104$\pm8 \degr$, we found that the resulting disc is roughly oriented in the 
same direction, although it rotates with a circular velocity
that is much higher.

\begin{figure}
   \includegraphics[width=\columnwidth]{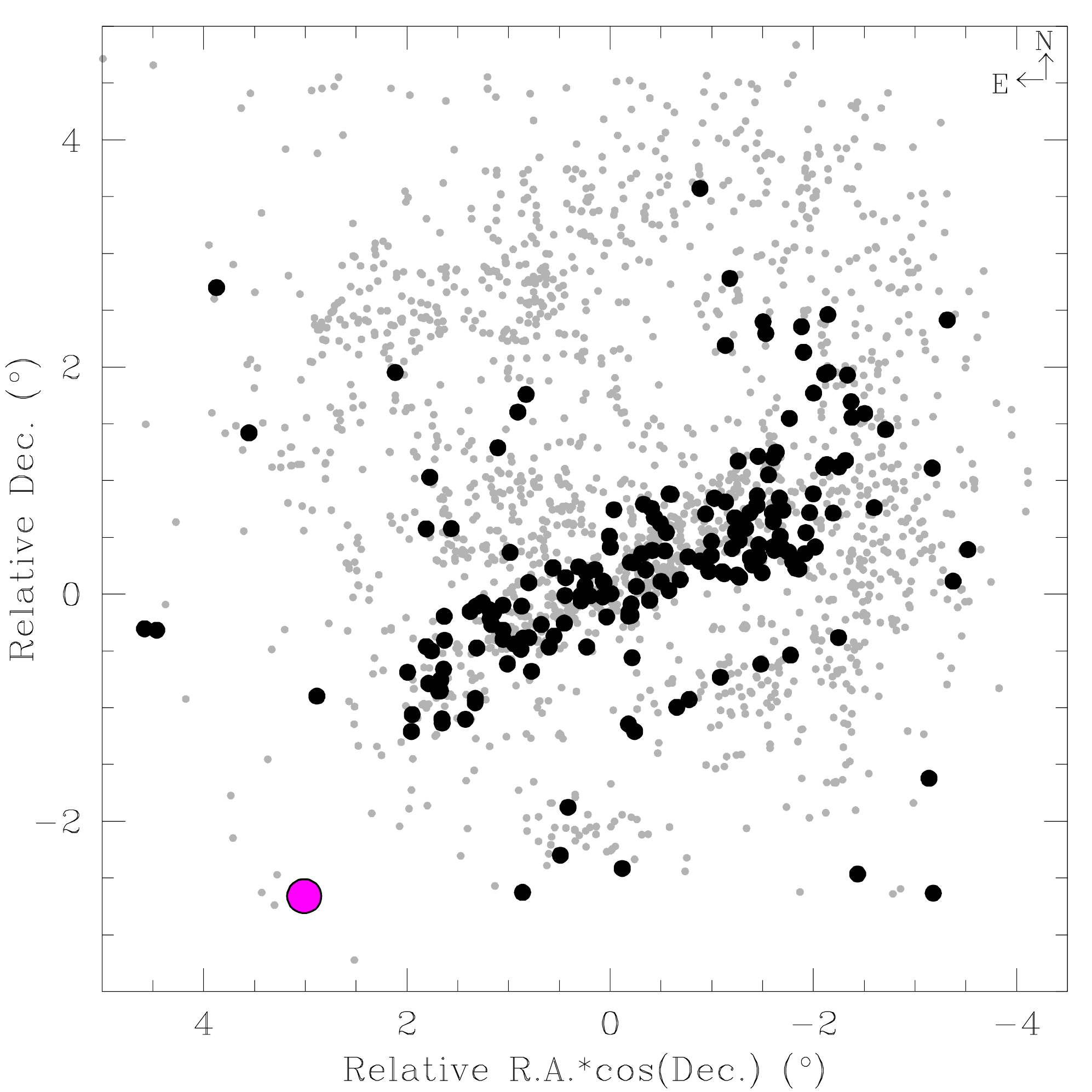}
    \caption{Spatial distribution of star clusters in \citet{betal08}'s catalogue
and those with age estimates  between log($t$ /yr) = 8.8 - 9.1 \citep{p14b} drawn with grey and black circles, respectively.
The new star cluster is represented by a magenta circle.}
   \label{fig:fig8}
\end{figure}

We speculate that such a higher circular velocity is a consequence of having been scattered
during an interaction with the MW/SMC from the inner LMC regions 
\citep{kallivayaliletal13,casettidinescuetal2014}, where it may have formed. 
Firstly, star clusters with age estimates within 
3$\times \sigma$ (age) from the cluster age, i.e. log($t$ /yr) between 8.8 and 9.1 not 
only are statistically more concentrated in the LMC bar, but also very
few star clusters have been identified in the outer Eastern side of the LMC disc. 
Therefore, it had to be scattered together with other star clusters of similar age
towards an outer disc-like orbit. Fig.~\ref{fig:fig8}
illustrates the spatial distribution of star clusters in the \citet{betal08} catalogue
with grey circles and those with ages between log($t$ /yr)$=$8.8 and 9.1 - taken from the
compilation of more than 2300 star clusters with age estimates \citep{p14b} - drawn
with black circles. The position in the sky of the new star cluster is represented
by the magenta circle. Secondly,  the metallicity level of
field stars in the outer LMC disc ($\rho >4 \degr$, $<$[Fe/H]$>$ = -0.9$\pm$0.2 dex) is on average
more metal-poor than that for inner disc field stars ($\rho <4 \degr$, $<$[Fe/H]$>$ =-0.5$\pm$0.2 dex)
\citep[see figure 3 in][]{pg13}, in excellent agreement with the observed outside-in galaxy chemical
enrichment \citep{hz09,carreraetal2011,meschin14}. Star clusters share the 
metallicities of their birthplaces. Nevertheless, with time ther drift away
from their birth locations. Interactions and other perturbations may
produce additional velocity components. Indeed, the derived cluster 
metallicity value ($<$[Fe/H]$>$=-0.45$\pm$0.15 
dex) is typical of the stellar population located within
 $\sim$ 2 degrees from the LMC centre, 
whereas the one expected for the new star cluster deprojected distance ($\rho=4.63\pm 0.03 \degr$) is in the outer
disc [Fe/H] range. This appears also to be the case, for instance, of other clusters 
younger than 2-3 Gyr with metallicities similar to that of the inner disc that are observed
in the outer disc \citep{getal06,pg13}.
Thirdly, the stellar density map of the new star 
cluster looks clearly elongated  along the SW-NE direction (see Fig.\ref{fig:fig5}), 
which may be a 
sign of tidal effects. Indeed, the position of the star cluster in the 
$r_c/r_h$ versus $r_h/rt$ plane suggests an object that has expanded up to its tidal radius 
\citep[][see, e.g., their figure 33.2]{hh03}, while its size is comparable to the
LMC globular clusters formed/steadily orbiting at deprojected distances $\la$ 1 degree;
globular clusters at $\rho \sim$ 5 degrees are in average 2-3 times bigger in radius
\citep{pm2018}. This means that the star cluster has the typical size of the innermost ones,
but lies in the outer LMC disc.  All of these arguments suggest that this cluster was born much
closer to the centre of the LMC and somehow was scattered to its current location.


\section{Conclusions}

We have carried out a photometric and spectroscopic study using
SOAR and Gemini observatory facilities of a recently new
discovered LMC star cluster, whose first derived astrophysical properties 
suggested that it was an object that was born in the inner LMC disc and soon
after ejected outwards into the intergalactic space, reaching a distance similar
to that of the SMC.

From accurate RV measurements of upper MS and red giant stars from the CaII
infrared triplet spectral region we assigned cluster memberships and estimated
the mean star cluster RV  and metallicity. By assuming a disc-like geometry, we found that the
circular velocity of a disc that rotates with the corresponding star cluster RV 
at the cluster deprojected distance is nearly 60 per cent higher than that
derived for most of the outer LMC disc clusters. The possibility that the
new star cluster has reached the outer LMC disc, after being born in the
innermost regions of the galaxy is supported by:

$\bullet$ the mean cluster distance (47.9 kpc) derived by fitting 
theoretical isochrone to the cluster CMD. The latter was built with stars 
distributed within the cluster radius, while the fit involved only two free 
parameters, namely: the cluster age and distance modulus, and relied on
the positions of RV member stars;

$\bullet$ the mean cluster age (0.9 Gyr), typical of star clusters mainly
formed along the LMC bar; the deprojected cluster distance from the LMC centre
 is $\rho= 4.6$ degrees (3.8 kpc).

$\bullet$ the mean cluster metallicity  ($<$[Fe/H]$>$=-0.45$\pm$0.15 dex) derived from
the CaII triplet line equivalent widths technique of red giant members.
This metal content is compatible with that from an outside-in formation scenario  
that brought more enriched gas out of which star clusters were born
towards the innermost regions;

$\bullet$ the cluster structural parameters derived from the fits of 
\citet{king62}, \citet{eff87} and \citet{plummer11} models to the
cluster stellar density radial profile extended from 0.1 out to 250 
arcsec from the cluster centre; the cluster radius is 31.6 arcsec
 (7.3 pc).

\section*{Acknowledgements}

Based on observations obtained at the Gemini Observatory, which is operated by 
the Association of Universities for Research in Astronomy, Inc., under a cooperative
 agreement with the NSF on behalf of the Gemini partnership: the National Science
 Foundation (United States), the National Research Council (Canada), CONICYT (Chile),
 Ministerio de Ciencia, Tecnolog\'{i}a e Innovaci\'{o}n Productiva (Argentina),
 and Minist\'{e}rio da Ci\^{e}ncia, Tecnologia e Inova\c{c}\~{a}o (Brazil). 
Based on observations obtained at the Southern Astrophysical Research (SOAR) telescope, 
which is a joint project of the Minist\'{e}rio da Ci\^{e}ncia, Tecnologia, 
Inova\c{c}\~{o}es e Comunica\c{c}\~{o}es (MCTIC) do Brasil, the U.S. National 
Optical Astronomy Observatory (NOAO), the University of North Carolina at Chapel 
Hill (UNC), and Michigan State University (MSU).
We thank the referee for the thorough reading of the manuscript and
timely suggestions to improve it. 




\bibliographystyle{mnras}

\begin{thebibliography}{}
\makeatletter
\relax
\def\mn@urlcharsother{\let\do\@makeother \do\$\do\&\do\#\do\^\do\_\do\%\do\~}
\def\mn@doi{\begingroup\mn@urlcharsother \@ifnextchar [ {\mn@doi@}
  {\mn@doi@[]}}
\def\mn@doi@[#1]#2{\def\@tempa{#1}\ifx\@tempa\@empty \href
  {http://dx.doi.org/#2} {doi:#2}\else \href {http://dx.doi.org/#2} {#1}\fi
  \endgroup}
\def\mn@eprint#1#2{\mn@eprint@#1:#2::\@nil}
\def\mn@eprint@arXiv#1{\href {http://arxiv.org/abs/#1} {{\tt arXiv:#1}}}
\def\mn@eprint@dblp#1{\href {http://dblp.uni-trier.de/rec/bibtex/#1.xml}
  {dblp:#1}}
\def\mn@eprint@#1:#2:#3:#4\@nil{\def\@tempa {#1}\def\@tempb {#2}\def\@tempc
  {#3}\ifx \@tempc \@empty \let \@tempc \@tempb \let \@tempb \@tempa \fi \ifx
  \@tempb \@empty \def\@tempb {arXiv}\fi \@ifundefined
  {mn@eprint@\@tempb}{\@tempb:\@tempc}{\expandafter \expandafter \csname
  mn@eprint@\@tempb\endcsname \expandafter{\@tempc}}}

\bibitem[\protect\citeauthoryear{{Belokurov} \& {Koposov}}{{Belokurov} \&
  {Koposov}}{2016}]{bk2016}
{Belokurov} V.,  {Koposov} S.~E.,  2016, \mn@doi [\mnras]
  {10.1093/mnras/stv2688}, \href
  {http://adsabs.harvard.edu/abs/2016MNRAS.456..602B} {456, 602}

\bibitem[\protect\citeauthoryear{{Bertin} \& {Arnouts}}{{Bertin} \&
  {Arnouts}}{1996}]{bertin96}
{Bertin} E.,  {Arnouts} S.,  1996, \mn@doi [\aaps] {10.1051/aas:1996164}, \href
  {http://adsabs.harvard.edu/abs/1996A%26AS..117..393B} {117, 393}

\bibitem[\protect\citeauthoryear{{Bica}, {Geisler}, {Dottori}, {Clari{\'a}},
  {Piatti}  \& {Santos}}{{Bica} et~al.}{1998}]{betal98}
{Bica} E.,  {Geisler} D.,  {Dottori} H.,  {Clari{\'a}} J.~J.,  {Piatti} A.~E.,
   {Santos} Jr. J.~F.~C.,  1998, \mn@doi [\aj] {10.1086/300448}, 116, 723

\bibitem[\protect\citeauthoryear{{Bica}, {Bonatto}, {Dutra}  \&
  {Santos}}{{Bica} et~al.}{2008}]{betal08}
{Bica} E.,  {Bonatto} C.,  {Dutra} C.~M.,   {Santos} J.~F.~C.,  2008, \mn@doi
  [\mnras] {10.1111/j.1365-2966.2008.13612.x}, 389, 678

\bibitem[\protect\citeauthoryear{{Bica}, {Santiago}, {Bonatto}, {Garcia-Dias},
  {Kerber}, {Dias}, {Barbuy}  \& {Balbinot}}{{Bica}
  et~al.}{2015}]{bicaetal2015}
{Bica} E.,  {Santiago} B.,  {Bonatto} C.,  {Garcia-Dias} R.,  {Kerber} L.,
  {Dias} B.,  {Barbuy} B.,   {Balbinot} E.,  2015, \mn@doi [\mnras]
  {10.1093/mnras/stv1720}, \href
  {http://adsabs.harvard.edu/abs/2015MNRAS.453.3190B} {453, 3190}

\bibitem[\protect\citeauthoryear{{Bressan}, {Marigo}, {Girardi}, {Salasnich},
  {Dal Cero}, {Rubele}  \& {Nanni}}{{Bressan} et~al.}{2012}]{betal12}
{Bressan} A.,  {Marigo} P.,  {Girardi} L.,  {Salasnich} B.,  {Dal Cero} C.,
  {Rubele} S.,   {Nanni} A.,  2012, \mn@doi [\mnras]
  {10.1111/j.1365-2966.2012.21948.x}, 427, 127

\bibitem[\protect\citeauthoryear{{Carrera}, {Gallart}, {Pancino}  \&
  {Zinn}}{{Carrera} et~al.}{2007}]{carreraetal2007}
{Carrera} R.,  {Gallart} C.,  {Pancino} E.,   {Zinn} R.,  2007, \mn@doi [\aj]
  {10.1086/520803}, \href {http://adsabs.harvard.edu/abs/2007AJ....134.1298C}
  {134, 1298}

\bibitem[\protect\citeauthoryear{{Carrera}, {Gallart}, {Aparicio}  \&
  {Hardy}}{{Carrera} et~al.}{2011}]{carreraetal2011}
{Carrera} R.,  {Gallart} C.,  {Aparicio} A.,   {Hardy} E.,  2011, \mn@doi [\aj]
  {10.1088/0004-6256/142/2/61}, \href
  {http://adsabs.harvard.edu/abs/2011AJ....142...61C} {142, 61}

\bibitem[\protect\citeauthoryear{{Casetti-Dinescu}, {Moni Bidin}, {Girard},
  {M{\'e}ndez}, {Vieira}, {Korchagin}  \& {van Altena}}{{Casetti-Dinescu}
  et~al.}{2014}]{casettidinescuetal2014}
{Casetti-Dinescu} D.~I.,  {Moni Bidin} C.,  {Girard} T.~M.,  {M{\'e}ndez}
  R.~A.,  {Vieira} K.,  {Korchagin} V.~I.,   {van Altena} W.~F.,  2014, \mn@doi
  [\apjl] {10.1088/2041-8205/784/2/L37}, \href
  {http://adsabs.harvard.edu/abs/2014ApJ...784L..37C} {784, L37}

\bibitem[\protect\citeauthoryear{{Cenarro}, {Cardiel}, {Gorgas}, {Peletier},
  {Vazdekis}  \& {Prada}}{{Cenarro} et~al.}{2001}]{cenarro01}
{Cenarro} A.~J.,  {Cardiel} N.,  {Gorgas} J.,  {Peletier} R.~F.,  {Vazdekis}
  A.,   {Prada} F.,  2001, \mn@doi [\mnras] {10.1046/j.1365-8711.2001.04688.x},
  \href {http://adsabs.harvard.edu/abs/2001MNRAS.326..959C} {326, 959}

\bibitem[\protect\citeauthoryear{{Cole}, {Smecker-Hane}, {Tolstoy}, {Bosler}
  \& {Gallagher}}{{Cole} et~al.}{2004}]{coleetal2004}
{Cole} A.~A.,  {Smecker-Hane} T.~A.,  {Tolstoy} E.,  {Bosler} T.~L.,
  {Gallagher} J.~S.,  2004, \mn@doi [\mnras]
  {10.1111/j.1365-2966.2004.07223.x}, \href
  {http://adsabs.harvard.edu/abs/2004MNRAS.347..367C} {347, 367}

\bibitem[\protect\citeauthoryear{{Cole}, {Tolstoy}, {Gallagher}  \&
  {Smecker-Hane}}{{Cole} et~al.}{2005}]{coleetal2005}
{Cole} A.~A.,  {Tolstoy} E.,  {Gallagher} III J.~S.,   {Smecker-Hane} T.~A.,
  2005, \mn@doi [\aj] {10.1086/428007}, \href
  {http://adsabs.harvard.edu/abs/2005AJ....129.1465C} {129, 1465}

\bibitem[\protect\citeauthoryear{{Da Costa}}{{Da Costa}}{2016}]{dacosta2016}
{Da Costa} G.~S.,  2016, \mn@doi [\mnras] {10.1093/mnras/stv2315}, \href
  {http://adsabs.harvard.edu/abs/2016MNRAS.455..199D} {455, 199}

\bibitem[\protect\citeauthoryear{{Elson}, {Fall}  \& {Freeman}}{{Elson}
  et~al.}{1987}]{eff87}
{Elson} R.~A.~W.,  {Fall} S.~M.,   {Freeman} K.~C.,  1987, \mn@doi [\apj]
  {10.1086/165807}, \href {http://adsabs.harvard.edu/abs/1987ApJ...323...54E}
  {323, 54}

\bibitem[\protect\citeauthoryear{{Feitzinger} \& {Weiss}}{{Feitzinger} \&
  {Weiss}}{1979}]{fw79}
{Feitzinger} J.~V.,  {Weiss} G.,  1979, \aaps, \href
  {http://adsabs.harvard.edu/abs/1979A%26AS...37..575F} {37, 575}

\bibitem[\protect\citeauthoryear{{Fraga}, {Kunder}  \& {Tokovinin}}{{Fraga}
  et~al.}{2013}]{fraga13}
{Fraga} L.,  {Kunder} A.,   {Tokovinin} A.,  2013, \mn@doi [\aj]
  {10.1088/0004-6256/145/6/165}, \href
  {http://adsabs.harvard.edu/abs/2013AJ....145..165F} {145, 165}

\bibitem[\protect\citeauthoryear{{Gimeno} et~al.,}{{Gimeno}
  et~al.}{2016}]{gimeno16}
{Gimeno} G.,  et~al., 2016, in Ground-based and Airborne Instrumentation for
  Astronomy VI. p. 99082S, \mn@doi{10.1117/12.2233883}

\bibitem[\protect\citeauthoryear{{Glatt}, {Grebel}  \& {Koch}}{{Glatt}
  et~al.}{2010}]{getal10}
{Glatt} K.,  {Grebel} E.~K.,   {Koch} A.,  2010, \mn@doi [\aap]
  {10.1051/0004-6361/201014187}, 517, A50

\bibitem[\protect\citeauthoryear{{Grocholski}, {Cole}, {Sarajedini}, {Geisler}
  \& {Smith}}{{Grocholski} et~al.}{2006}]{getal06}
{Grocholski} A.~J.,  {Cole} A.~A.,  {Sarajedini} A.,  {Geisler} D.,   {Smith}
  V.~V.,  2006, \mn@doi [\aj] {10.1086/507303}, 132, 1630

\bibitem[\protect\citeauthoryear{{Hammer}, {Yang}, {Flores}, {Puech}  \&
  {Fouquet}}{{Hammer} et~al.}{2015}]{hammeretal2015}
{Hammer} F.,  {Yang} Y.~B.,  {Flores} H.,  {Puech} M.,   {Fouquet} S.,  2015,
  \mn@doi [\apj] {10.1088/0004-637X/813/2/110}, \href
  {/abs/2015ApJ...813..110H} {813, 110}

\bibitem[\protect\citeauthoryear{{Hanuschik}}{{Hanuschik}}{2003}]{hanuschik03}
{Hanuschik} R.~W.,  2003, \mn@doi [\aap] {10.1051/0004-6361:20030885}, \href
  {http://adsabs.harvard.edu/abs/2003A%26A...407.1157H} {407, 1157}

\bibitem[\protect\citeauthoryear{{Harris} \& {Zaritsky}}{{Harris} \&
  {Zaritsky}}{2009}]{hz09}
{Harris} J.,  {Zaritsky} D.,  2009, \mn@doi [\aj]
  {10.1088/0004-6256/138/5/1243}, 138, 1243

\bibitem[\protect\citeauthoryear{{Heggie} \& {Hut}}{{Heggie} \&
  {Hut}}{2003}]{hh03}
{Heggie} D.,  {Hut} P.,  2003, {The Gravitational Million-Body Problem: A
  Multidisciplinary Approach to Star Cluster Dynamics}

\bibitem[\protect\citeauthoryear{{Hook}, {J{\o}rgensen}, {Allington-Smith},
  {Davies}, {Metcalfe}, {Murowinski}  \& {Crampton}}{{Hook}
  et~al.}{2004}]{hook04}
{Hook} I.~M.,  {J{\o}rgensen} I.,  {Allington-Smith} J.~R.,  {Davies} R.~L.,
  {Metcalfe} N.,  {Murowinski} R.~G.,   {Crampton} D.,  2004, \mn@doi [\pasp]
  {10.1086/383624}, \href {http://adsabs.harvard.edu/abs/2004PASP..116..425H}
  {116, 425}

\bibitem[\protect\citeauthoryear{{Indu} \& {Subramaniam}}{{Indu} \&
  {Subramaniam}}{2015}]{is2015}
{Indu} G.,  {Subramaniam} A.,  2015, \mn@doi [\aap]
  {10.1051/0004-6361/201321133}, \href
  {http://adsabs.harvard.edu/abs/2015A%26A...573A.136I} {573, A136}

\bibitem[\protect\citeauthoryear{{Kallivayalil}, {van der Marel}, {Besla},
  {Anderson}  \& {Alcock}}{{Kallivayalil} et~al.}{2013}]{kallivayaliletal13}
{Kallivayalil} N.,  {van der Marel} R.~P.,  {Besla} G.,  {Anderson} J.,
  {Alcock} C.,  2013, \mn@doi [\apj] {10.1088/0004-637X/764/2/161}, \href
  {http://adsabs.harvard.edu/abs/2013ApJ...764..161K} {764, 161}

\bibitem[\protect\citeauthoryear{{King}}{{King}}{1962}]{king62}
{King} I.,  1962, \mn@doi [\aj] {10.1086/108756}, 67, 471

\bibitem[\protect\citeauthoryear{{Mackey}, {Koposov}, {Erkal}, {Belokurov}, {Da
  Costa}  \& {G{\'o}mez}}{{Mackey} et~al.}{2016}]{mackeyetal2016}
{Mackey} A.~D.,  {Koposov} S.~E.,  {Erkal} D.,  {Belokurov} V.,  {Da Costa}
  G.~S.,   {G{\'o}mez} F.~A.,  2016, \mn@doi [\mnras] {10.1093/mnras/stw497},
  \href {http://adsabs.harvard.edu/abs/2016MNRAS.459..239M} {459, 239}

\bibitem[\protect\citeauthoryear{{McLaughlin}, {Harris}  \&
  {Hanes}}{{McLaughlin} et~al.}{1994}]{mclaughlin94}
{McLaughlin} D.~E.,  {Harris} W.~E.,   {Hanes} D.~A.,  1994, \mn@doi [\apj]
  {10.1086/173744}, \href {http://adsabs.harvard.edu/abs/1994ApJ...422..486M}
  {422, 486}

\bibitem[\protect\citeauthoryear{{Meschin}, {Gallart}, {Aparicio}, {Hidalgo},
  {Monelli}, {Stetson}  \& {Carrera}}{{Meschin} et~al.}{2014}]{meschin14}
{Meschin} I.,  {Gallart} C.,  {Aparicio} A.,  {Hidalgo} S.~L.,  {Monelli} M.,
  {Stetson} P.~B.,   {Carrera} R.,  2014, \mn@doi [\mnras]
  {10.1093/mnras/stt2220}, 438, 1067

\bibitem[\protect\citeauthoryear{{Moni Bidin}, {Casetti-Dinescu}, {Girard},
  {Zhang}, {M{\'e}ndez}, {Vieira}, {Korchagin}  \& {van Altena}}{{Moni Bidin}
  et~al.}{2017}]{mbetal2017}
{Moni Bidin} C.,  {Casetti-Dinescu} D.~I.,  {Girard} T.~M.,  {Zhang} L.,
  {M{\'e}ndez} R.~A.,  {Vieira} K.,  {Korchagin} V.~I.,   {van Altena} W.~F.,
  2017, \mn@doi [\mnras] {10.1093/mnras/stw3242}, \href
  {http://adsabs.harvard.edu/abs/2017MNRAS.466.3077M} {466, 3077}

\bibitem[\protect\citeauthoryear{{Piatti}}{{Piatti}}{2014}]{p14b}
{Piatti} A.~E.,  2014, \mn@doi [\mnras] {10.1093/mnras/stt1998}, 437, 1646

\bibitem[\protect\citeauthoryear{{Piatti}}{{Piatti}}{2016}]{p16}
{Piatti} A.~E.,  2016, \mn@doi [\mnras] {10.1093/mnrasl/slw053}, \href
  {http://adsabs.harvard.edu/abs/2016MNRAS.459L..61P} {459, L61}

\bibitem[\protect\citeauthoryear{{Piatti}}{{Piatti}}{2017}]{p17a}
{Piatti} A.~E.,  2017, \mn@doi [\apjl] {10.3847/2041-8213/834/2/L14}, \href
  {http://adsabs.harvard.edu/abs/2017ApJ...834L..14P} {834, L14}

\bibitem[\protect\citeauthoryear{{Piatti} \& {Geisler}}{{Piatti} \&
  {Geisler}}{2013}]{pg13}
{Piatti} A.~E.,  {Geisler} D.,  2013, \mn@doi [\aj]
  {10.1088/0004-6256/145/1/17}, 145, 17

\bibitem[\protect\citeauthoryear{{Piatti} \& {Mackey}}{{Piatti} \&
  {Mackey}}{2018}]{pm2018}
{Piatti} A.~E.,  {Mackey} A.~D.,  2018, \mn@doi [\mnras (in press)]
  {10.1093/mnras/sty1048}, \href
  {http://adsabs.harvard.edu/abs/2018MNRAS.tmp..991P} {}

\bibitem[\protect\citeauthoryear{{Piatti}, {Geisler}, {Sarajedini}  \&
  {Gallart}}{{Piatti} et~al.}{2009}]{piattietal2009}
{Piatti} A.~E.,  {Geisler} D.,  {Sarajedini} A.,   {Gallart} C.,  2009, \mn@doi
  [\aap] {10.1051/0004-6361/200912223}, \href
  {http://adsabs.harvard.edu/abs/2009A%26A...501..585P} {501, 585}

\bibitem[\protect\citeauthoryear{{Piatti}, {de Grijs}, {Rubele}, {Cioni},
  {Ripepi}  \& {Kerber}}{{Piatti} et~al.}{2015}]{petal15a}
{Piatti} A.~E.,  {de Grijs} R.,  {Rubele} S.,  {Cioni} M.-R.~L.,  {Ripepi} V.,
   {Kerber} L.,  2015, \mn@doi [\mnras] {10.1093/mnras/stv635}, 450, 552

\bibitem[\protect\citeauthoryear{{Piatti}, {Hwang}, {Cole}, {Angelo}  \&
  {Emptage}}{{Piatti} et~al.}{2018}]{piattietal2018}
{Piatti} A.~E.,  {Hwang} N.,  {Cole} A.~A.,  {Angelo} M.~S.,   {Emptage} B.,
  2018, \mn@doi [\mnras] {10.1093/mnras/sty2324}, \href
  {http://adsabs.harvard.edu/abs/2018MNRAS.tmp.2198P} {}

\bibitem[\protect\citeauthoryear{{Pieres} et~al.,}{{Pieres}
  et~al.}{2016}]{pieresetal2016}
{Pieres} A.,  et~al., 2016, \mn@doi [\mnras] {10.1093/mnras/stw1260}, \href
  {http://adsabs.harvard.edu/abs/2016MNRAS.461..519P} {461, 519}

\bibitem[\protect\citeauthoryear{{Plummer}}{{Plummer}}{1911}]{plummer11}
{Plummer} H.~C.,  1911, \mn@doi [\mnras] {10.1093/mnras/71.5.460}, \href
  {http://adsabs.harvard.edu/abs/1911MNRAS..71..460P} {71, 460}

\bibitem[\protect\citeauthoryear{{Salem}, {Besla}, {Bryan}, {Putman}, {van der
  Marel}  \& {Tonnesen}}{{Salem} et~al.}{2015}]{salemetal2015}
{Salem} M.,  {Besla} G.,  {Bryan} G.,  {Putman} M.,  {van der Marel} R.~P.,
  {Tonnesen} S.,  2015, \mn@doi [\apj] {10.1088/0004-637X/815/1/77}, \href
  {http://adsabs.harvard.edu/abs/2015ApJ...815...77S} {815, 77}

\bibitem[\protect\citeauthoryear{{Salinas}, {Alabi}, {Richtler}  \&
  {Lane}}{{Salinas} et~al.}{2015}]{salinas15}
{Salinas} R.,  {Alabi} A.,  {Richtler} T.,   {Lane} R.~R.,  2015, \mn@doi
  [\aap] {10.1051/0004-6361/201425574}, \href
  {http://adsabs.harvard.edu/abs/2015A%26A...577A..59S} {577, A59}

\bibitem[\protect\citeauthoryear{{Salinas}, {Contreras Ramos}, {Strader},
  {Hakala}, {Catelan}, {Peacock}  \& {Simunovic}}{{Salinas}
  et~al.}{2016}]{salinas16}
{Salinas} R.,  {Contreras Ramos} R.,  {Strader} J.,  {Hakala} P.,  {Catelan}
  M.,  {Peacock} M.~B.,   {Simunovic} M.,  2016, \mn@doi [\aj]
  {10.3847/0004-6256/152/3/55}, \href
  {http://adsabs.harvard.edu/abs/2016AJ....152...55S} {152, 55}

\bibitem[\protect\citeauthoryear{{Schirmer}}{{Schirmer}}{2013}]{schirmer13}
{Schirmer} M.,  2013, \mn@doi [\apjs] {10.1088/0067-0049/209/2/21}, \href
  {http://adsabs.harvard.edu/abs/2013ApJS..209...21S} {209, 21}

\bibitem[\protect\citeauthoryear{{Schommer}, {Suntzeff}, {Olszewski}  \&
  {Harris}}{{Schommer} et~al.}{1992}]{s92}
{Schommer} R.~A.,  {Suntzeff} N.~B.,  {Olszewski} E.~W.,   {Harris} H.~C.,
  1992, \mn@doi [\aj] {10.1086/116074}, \href
  {http://adsabs.harvard.edu/abs/1992AJ....103..447S} {103, 447}

\bibitem[\protect\citeauthoryear{{Sharma}, {Borissova}, {Kurtev}, {Ivanov}  \&
  {Geisler}}{{Sharma} et~al.}{2010}]{shetal10}
{Sharma} S.,  {Borissova} J.,  {Kurtev} R.,  {Ivanov} V.~D.,   {Geisler} D.,
  2010, \mn@doi [\aj] {10.1088/0004-6256/139/3/878}, 139, 878

\bibitem[\protect\citeauthoryear{{Sitek} et~al.,}{{Sitek}
  et~al.}{2016}]{siteketal2016}
{Sitek} M.,  et~al., 2016, \actaa, \href
  {http://adsabs.harvard.edu/abs/2016AcA....66..255S} {66, 255}

\bibitem[\protect\citeauthoryear{{Stetson}}{{Stetson}}{1987}]{stetson87}
{Stetson} P.~B.,  1987, \mn@doi [\pasp] {10.1086/131977}, \href
  {http://adsabs.harvard.edu/abs/1987PASP...99..191S} {99, 191}

\bibitem[\protect\citeauthoryear{{Stetson}}{{Stetson}}{1992}]{stetson92}
{Stetson} P.~B.,  1992, \jrasc, \href
  {http://adsabs.harvard.edu/abs/1992JRASC..86...71S} {86, 71}

\bibitem[\protect\citeauthoryear{{Taylor}}{{Taylor}}{2006}]{taylor06}
{Taylor} M.~B.,  2006, in {Gabriel} C.,  {Arviset} C.,  {Ponz} D.,   {Enrique}
  S.,  eds,  Astronomical Society of the Pacific Conference Series Vol. 351,
  Astronomical Data Analysis Software and Systems XV. p.~666

\bibitem[\protect\citeauthoryear{{Tokovinin}, {Cantarutti}, {Tighe},
  {Schurter}, {Martinez}, {Thomas}  \& {van der Bliek}}{{Tokovinin}
  et~al.}{2016}]{tokovinin16}
{Tokovinin} A.,  {Cantarutti} R.,  {Tighe} R.,  {Schurter} P.,  {Martinez} M.,
  {Thomas} S.,   {van der Bliek} N.,  2016, \mn@doi [\pasp]
  {10.1088/1538-3873/128/970/125003}, \href
  {http://adsabs.harvard.edu/abs/2016PASP..128l5003T} {128, 125003}

\bibitem[\protect\citeauthoryear{{Tonry} \& {Davis}}{{Tonry} \&
  {Davis}}{1979}]{tonry79}
{Tonry} J.,  {Davis} M.,  1979, \mn@doi [\aj] {10.1086/112569}, \href
  {http://adsabs.harvard.edu/abs/1979AJ.....84.1511T} {84, 1511}

\bibitem[\protect\citeauthoryear{{van der Marel} \& {Kallivayalil}}{{van der
  Marel} \& {Kallivayalil}}{2014}]{vdmk14}
{van der Marel} R.~P.,  {Kallivayalil} N.,  2014, \mn@doi [\apj]
  {10.1088/0004-637X/781/2/121}, 781, 121

\bibitem[\protect\citeauthoryear{{van der Marel}, {Alves}, {Hardy}  \&
  {Suntzeff}}{{van der Marel} et~al.}{2002}]{vdmareletal2002}
{van der Marel} R.~P.,  {Alves} D.~R.,  {Hardy} E.,   {Suntzeff} N.~B.,  2002,
  \mn@doi [\aj] {10.1086/343775}, \href
  {http://adsabs.harvard.edu/abs/2002AJ....124.2639V} {124, 2639}

\makeatother
\end{thebibliography}

\input{paper.bbl}







\bsp	
\label{lastpage}
\end{document}